# Unveiling the physics of the Thomson jumping ring

**Celso L. Ladera and Guillermo Donoso**

**Departamento de Física, Universidad Simón Bolívar**

**Caracas 1086, Venezuela**

*clladera@usb.ve*

We present a new model, and the validating experiments, that unveil the rich physics behind the flight of a conductive ring in the Thomson experiment, a physics veiled by the fast thrust that impels the ring. We uncover interesting features of the electro-dynamics of the flying ring, e.g. the varying mutual inductance between ring and the thrusting electromagnet, or how to measure the ring proper magnetic field in the presence of the larger field of the electromagnet. We succeed in separating the position and time dependences of the ring variables as it travels upward in a diverging magnetic field, obtaining a comprehensive view of the ring motion. We introduce a low-cost jumping ring set-up that incorporates simple innovative devices, e.g. a couple of pick-up coils connected in opposition that allows us to scrutinize the ring electro-dynamics, and to confirm the predictions of our theoretical model with good accuracy. This work is within the reach of senior students of science or engineering, and it can be exploited either as a teaching laboratory experiment or as an open-end project work.

## I. INTRODUCTION

Many of the published works on the Thomson jumping ring experiment are naturally biased towards showing how high an electro-magnet is capable of throwing a conductive ring when excited with a current. The latter was indeed the original purpose of E. Thomson in the 1880's when it was crucially important to show that AC currents could do work, in the debate on DC or AC public networks. And no doubt, in demonstrations of Thomson ring experiment what catches more attention, from students or public, is the sight of ring flying upward, as high as possible, suddenly thrust by the electro-magnet. That thrust happens fast and little attention is paid to the plethora of physics facts behind the experiment: as frequently heard "it all can be explained by just applying Faraday´s Law and Lenz law". In some instances a large charged capacitor is incorporated in the circuit of the thrusting device, which once fast discharged as a current pulse through the electro-magnet propels the ring to several meters high. Nonetheless, the last two decades have seen interesting works published on the subject, in which more details of the physics behind Thomson experiment has become to be unveiled [1-5]. For instance, the seldom demonstrated levitation of the ring on the electromagnet, and the time the ring takes in abandoning the electromagnet and the dependence of the force on the ring upon the frequency of the alternating current, are now receiving attention. We recently published [5] a work in which the explicit functional dependences of the ring motion on its most relevant variables were theoretically and experimentally found and demonstrated, some of them for the first time. For instance we showed how to measure the instantaneous current induced in the ring itself, by the electro-magnet, or measured the magnetic force on the ring as a function of the excitation current for increasing time-frequencies of that current. In this work we focus our attention in the flight of the ring, to unveil both theoretically and experimentally, and in detailed ways, the explicit dependences of that flight on variables such as: the number of cycles of the applied AC excitation to the electro-magnet in a given thrust, the varying current that appears in the electromagnet coil ant its relation to the mutual inductance between ring and electro-magnet, the ring velocity and acceleration functions, the force on the ring as a function of height, and even the ring height dependence of the actual electro-motive force (*e.m.f.*) induced in the proper ring. Our experiments are done using a modest-size jumping ring apparatus and our ring only reaches several tens of centimeters high, with modest coil currents, but our results and conclusions are indeed valid for the more massive version of the AC driven experiment. Moreover, our apparatus incorporates devices not used before to scrutinize the ring motion and its physics in a new way, and thus we get true experimental results not published before. Two long vertical pick-up coils wound over the electro-magnet allow us to measure variables such as the actual current induced in the ring, or the varying mutual inductance of the transformer-like system *jumping ring-electromagnet*, and a LED illuminated slit-source and phototransistor device allows us to accurately monitor the ring position along its vertical flight. We also develop here a new analytical approach in which the dependences of the ring electro-dynamics upon time and position are conveniently separated, determined, and then successfully validated with accuracy and precision by the set of experiments presented in Section IV. Both, our theoretical model and our experiments – done with a minimum of instrumentation – are within the reach of undergraduate students of science or engineering. This work is appropriate for implementation as a senior physics laboratory experiment, and appropriate indeed for an open-end project work of low-cost.

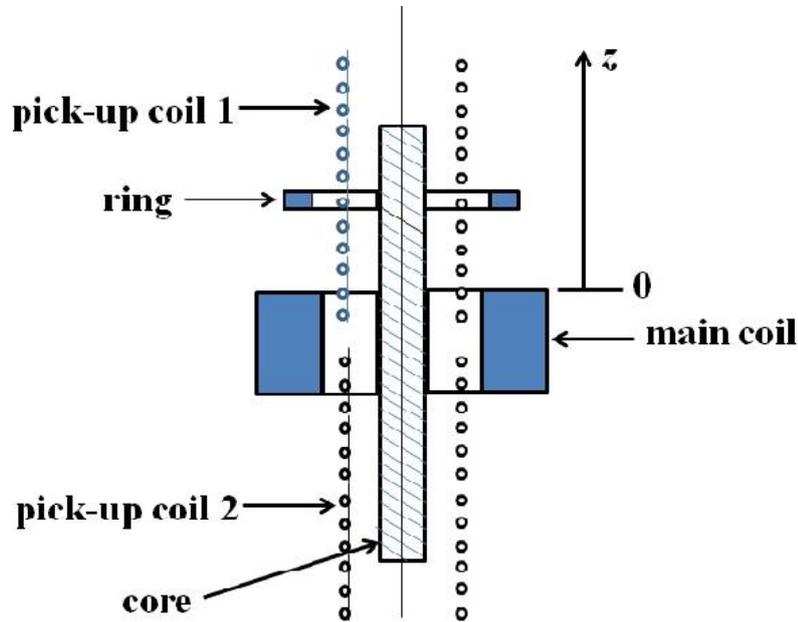

**Figure 1** Jumping ring set-up (longitudinal section) showing the coil, the long core of the electromagnet an the ring. Two long pick-up coils 1 and 2 connected in opposition allow the measurement of both the electrical current and the electro-motive force induced in the ring. Not shown, we attach a vertical ruled transparent strip to the ring rim that allows us to directly monitor the position of the ring as it flies upward.

## II. A DETAILED ANALYTICAL MODEL FOR THE JUMPING RING

Consider the jumping ring set-up illustrated in Fig. 1. The conductive ring is coaxial with an electro-magnet whose coil (also called *main coil* below) is excited with an AC current of known angular frequency $\omega$. We are to unveil the whole motion of the flying ring along the vertical *z*-axis by solving its electro-dynamics motion equation. This is not straightforward to do because as the reader soon discovers we have to deal with a case in which the ring motion takes place in a time-varying non-uniform magnetic field created by an electro-magnet whose electrical parameters happen to be dependent on the motion of the ring itself! Thus, we are interested in a detailed account of the force on the ring (assumed of mean radius *a*) as it flies upward. The instantaneous value of this magnetic force is given by the well-known physics textbooks relation [6, 7] that accounts for the force exerted by a magnetic field on a ring current

$$F(t) = i_{ring}(t)\, 2\pi a\, B_\rho(t), \qquad (1)$$

where $i_{ring}$ is the time-varying current induced in the ring by the time-varying magnetic field of the electro-magnet. Here $B_\rho(t)$ represents the harmonically varying radial component (of amplitude $B_{\rho,m}$) of that field i.e.

$$B_\rho(t) = B_{\rho,m}\, \sin \omega t \qquad (2)$$

for an exciting current $I_c(t) = I_m \sin \omega t$ present in the main coil. The instantaneous electrical current in the ring is given by the quotient of its induced electro-motive force $\varepsilon_i(t)$ and its electrical impedance $Z(z)$. The first one can be obtained by applying Faraday´s Law to the time-varying axial magnetic flux of the core (assumed to be of squared cross-section of area $b^2$), actually to the flux of the vertical component $B_z(t)$ of the main coil field, i.e.

$$\varepsilon_i(t) = -\frac{d}{dt}[b^2 B_z(t)] = -\omega\, b^2 B_{z,m}\, \cos \omega t \qquad (3)$$

while the second $Z(z)$ is not constant but a function of the ring vertical position $z$ as the latter flies upward, and it is given by $Z = \sqrt{R^2 + \omega^2 L(z)^2}$, where $R$ and $L(z)$ denote the resistance and inductance of the moving ring, respectively. As seen in Fig. 1 the origin $z=0$ of the vertical coordinate axis we fix at the top of the electromagnet (where the ring rests when no electrical current is applied to the electromagnet). The vertical position varying inductance of the ring shall be conveniently written later on as $L(z) = l(z)\, L_0$, where $L_0$ is the nominal ring self-inductance (a purely geometry dependent parameter), and $l(z)$ is a functional factor that accounts for the vertical position dependence of the ring inductance. It is important to be aware and recall that as the ring travels upward it does so across the non-uniform magnetic field region above the electromagnet: There the magnetic lines are divergent, and their density diminishes as the ring vertical position increases (the field weakens both radially and axially). Therefore, all the relevant electro-dynamics variables of the ring vary with vertical position and time.

We found it to be convenient to separate the functional dependences of the axial component of the magnetic field upon time $t$ and vertical coordinate $z$ as follows,

$$B_z(t) = B_{1z}\, b_z(z)\, I_c(z,t), \qquad (4)$$

or
$$B_z(t) = B_{1z} \, b_z(z) \, I_c(t) I(z) \tag{5}$$

where $B_{1z}$ is the constant value of the axial field component at the top of the electromagnet core (i.e. at $z=0$), and generated by a coil excitation current $I_c$ of exactly 1.0 A. Here $b_z(z)$ is an auxiliary $z$-functional factor that represents the axial variation of the magnetic field produced by the magnetic permeability of the electromagnet core. This $b_z(z)$ can be readily determined experimentally, and should be normalized so that $b_z(0)=1$. Moreover, $I(z)$ is the auxiliary functional factor that incorporates into our model the decrease produced on the coil exciting current by the vertically varying mutual inductance between the flying ring and the coil. It is too an auxiliary function to be determined as shall be explained below (Appendix A). The two components of the electromagnet field can also be theoretically obtained, with some extra effort (because of the electromagnet core) resorting to the treatment of coil fields presented in [8].

Analogously, the radial component of the magnetic field may be written as

$$B_\rho(t) = B_{1\rho} \, b_\rho(z) \, I_c(t) I(z), \tag{6}$$

for the coil current $I_c(t) = I_m \sin \omega t$. We may then rewrite the two field components as

$$B_z(z,t) = B_{1z} \, b_z(z) \, I_m \, I(z) \sin \omega t \tag{7}$$

$$B_\rho(z,t) = B_{1\rho} \, b_\rho(z) \, I_m \, I(z) \sin \omega t. \tag{8}$$

The required *e.m.f.* induced in the ring can now be written as,

$$\varepsilon_i(z,t) = \omega \, b^2 B_{1z} \, b_z(z) \, I_c(0) I(z) \cos \omega t = \varepsilon_{i,max} \cos \omega t, \tag{9}$$

and then the current in the travelling ring becomes

$$i_{ring}(z,t) = \frac{\varepsilon_{i,max}}{Z} \cos[\omega t - \phi(z)] = \frac{\varepsilon_{i,max}}{\omega L} \sin \phi(z) \cos[\omega t - \phi(z)]. \tag{10}$$

Here $\phi(z)$, such that $\tan \phi(z) = \frac{\omega L(z)}{R}$, is the expected ring position-dependent phase difference between the current in the ring and the *e.m.f.* induced in the ring itself, by the driving magnetic field applied by the electromagnet. We are now ready to write the instantaneous magnetic force $F(z,t)$ in the ring (after Eq. (1)) in a convenient way,

$$F(z,t) = \frac{1}{L(z)} (2\pi a)\, b^2\, B_{1z}\, b_z(z) B_{1\rho} b_\rho(z) \sin\phi(z) \times \ldots$$

$$\cos[\omega t - \phi(z)]\, I_c(0)^2 I(z)^2 \sin\omega t \qquad (11)$$

Note that this is an oscillating force at twice the frequency of the AC voltage applied to the electromagnet to impulse the ring. Although a bit long the latter expression will allows one to determine the ring electrodynamics at any height and time (note for instance the presence in the equation of the vertically varying inductance $L(z)$ which controls the coupling between ring and the magnetic field gradient). An interesting variable, often mentioned in the literature is the *time average force F(z)* on the magnet, still a function of the ring vertical position $z$. This average we can readily obtain from the equation above, and in such a way that we can envisage the separate influences of the axial or the radial magnetic field components. In our case the average force on the ring is given (from Eq. (11)) by

$$F(z) = \frac{1}{l(z) L_0} (\pi a b^2)\, B_{1z}\, b_z(z) B_{1\rho} b_\rho(z) I_c(0)^2 I(z)^2 \sin^2\phi(z) \qquad (12)$$

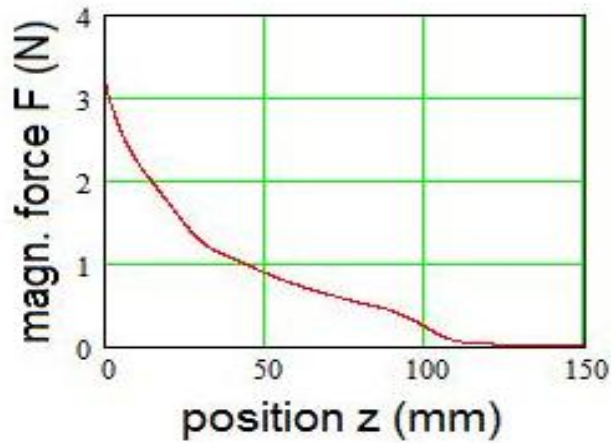

**Fig. 2** Predicted magnetic force in N (see Eq. 12) on the ring as a function of the ring position. The force decays monotonically to zero at about 11 cm (in coincidence with the upper end of the protruding core shown in Fig. 1).

Both, the calculation of the instantaneous force *F(z,t)* and its time-average demand from us to determine the three *position-dependent* auxiliary functions introduced before, namely $b_\rho(z)$ for the field radial component, $b_z(z)$ for the axial field component,, and $l(z)$ for the varying inductance. These functions fortunately we can find with little effort, and will be dealt with later on in this work.

Having found an expression for the magnetic force on the flying ring (of known mass *m*), we may now consider solving its motion differential equation which is of course

$$\frac{d^2}{dt^2} z(t) = F[z(t)] - mg \qquad (13)$$

After introducing the expression for the force *F(z)* found above, this equation can be solved using a numerical procedure e.g. the second-order Runge-Kutta method (or resorting to any well-known mathematics commercial software packages) with the initial conditions *z(0)=0* and *z'(0)=0*. A couple of typical solutions for the ring position *z(t)* and its speed *z´(t)* shall be plotted below (see Section II) where they are to be compared with our laboratory measured vertical position and speed of the ring as it travels upward.

## III. APPARATUS AND EXPERIMENTAL SET-UP

Our experimental set-up is similar to the one seen in standard classroom jumping ring experiments: an electromagnet vertically aligned and a coaxial metallic ring, plus an AC constant voltage source. The electromagnet we use has a 30 cm long squared-section steel core of side *b = 4 cm*. Its 6.5 cm long coil has 500 turns of thick copper wire (AWG 18, 1 mm diameter), and its nominal resistance and inductance are 2.5 Ω and 11 mH, respectively. It is driven using a Variac at the mains frequency of 60 Hz. In our experiments this coil carries currents of amplitudes $I_c(0)$ in the range 1-10 A. Our typical ring, made of aluminum stock, has a mean radius of 3.65 cm and a mass of 68.3 g. The initial position *(z=0)* of our ring is at the top of the coil (Fig. 1). The resistance of the ring is a mere *0.070 mΩ*, and its nominal inductance is rather small too *$L_0$=0.110 µH*. A digital dual-trace oscilloscope is used to monitor the signals in the electro-magnet coil and in the pick-up coils shown in Fig. 1. Our magnetometer is a commercial *x-y-z* magnetic field Hall probe [9].

But, we have incorporated in our set-up additional devices in order to obtain detailed experimental evidence of the ring electrodynamics. Thus, along the whole electromagnet core there are two thin-wire (AWG 38, 0.1007 mm diameter) elongated pick-up coils 19 cm long each one that are hand-wound, at two turns per centimeter, on a rolled sheet of bond paper. These coils surround the vertical core (they go in the small space left between main coil and core). These two pick-up coils are symmetrically located and connected in opposition at the middle so that their signals are *π-rad* out of phase. The upper pick-up coil spans from the mid plane of the core to about 4 cm above the top of the core, and covers the region where the ring interacts with electromagnet field and produces its own magnetic field. The lower one spans from the middle of the core to 4 cm below the bottom of the core, far from where the ring moves, and therefore its interaction with the magnetic field produced by the ring itself is practically nil. Connected in opposition the two pick-up coils cancel the magnetic field effects of the main coil (we measured a 500/2 ratio of induced electrical signals between these two pick-up coils) and allows us to measure the magnetic field produced by the induced current in the ring in the presence of the main coil field. A digital dual-trace oscilloscope was used to monitor the signals of the electro-magnet coil and of the pick-up coils.

To accurately monitor the vertical position of the ring as it travels upward we attached to it a long strip of transparent thin acrylic sheet a scale of 2 mm divisions along the first centimeter, and every 5 mm afterwards, ruled out on it. This light strip hangs down from a side of the ring. A horizontal thin collimated beam of light from a narrow slit illuminated from behind with a white LED, traverses the strip to fall on a photo-transistor followed by a diode connected to the phototransistor emitter. As the ring moves upward we can then register its position by monitoring the signal form this simple circuit with a digital oscilloscope. This position monitoring system works very well: it gives us the ring vertical position with a resolution of about 0.5 mm (see also Fig. 5 below).

We have also resorted to an auxiliary little pick-up coil, 10 turns of thin wire (38 AWG), 36.5 mm mean radius and only 1 mm high, to measure the vertical component of the electromagnet field along the vertical axis, by simply step-sliding it up and measuring the field through it.

## IV.    EXPERIMENTS AND RESULTS

### (a) Exciting current in the coil and the *e.mf.* induced in the pick-up coils by the ring itself

Figure 3 shows actual oscilloscope traces of the exciting current in the main coil of the electromagnet (upper trace) and of the *e.m.f.* induced in the long pick-up coils by the magnetic field generated by the ring itself (lower trace). The main coil trace corresponds to 13 cycles of 80 V constant amplitude voltage applied to the coil with the Variac (80 V amplitude corresponds to an excitation current of about 5 A amplitude).

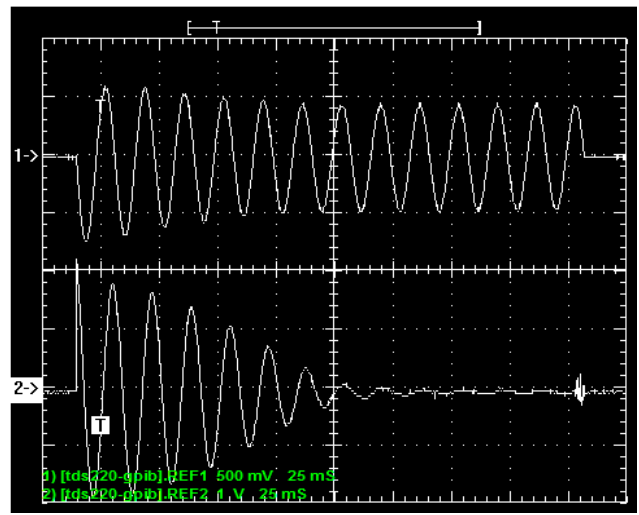

**Fig. 3** Oscilloscope traces of the electric current in the main coil and the *e.m.f.* induced in the pick-up coils by the magnetic field produced by the jumping ring (for 13 cycles of 80 V of constant amplitude, 60 Hz voltage applied to the electro-magnet)

These two traces contain very useful information. For instance, note that in spite of the constant voltage applied to the coil the amplitude of the current $I_c = I_m\, I(z(t))\, \sin(\omega t)$ in the coil decreases with time. The decay of this current we attribute to the ring as it moves upward. In effect, the whole apparatus we may consider as a transformer in which the main coil is the *primary* while the ring plays the role of the *secondary*. The effective inductance of the coil $L´$ depends upon the mutual inductance $M$ between the ring and the coil (see Appendix A). This coupling inductance $M$ diminishes as the ring moves upward, away from the electromagnet, which means that $L´$ increases and then the coil impedance increases too. In summary the coil current must diminish as the ring withdraws upward from the coil. The pick-up coils signal (lower trace of Fig. 3) is closely related to the electrical current that appears in the ring as we explain in the sub-section IV (b) below.

There is a phase difference $\phi$ between the two traces in Fig. 3 which can be directly measured by simply superposing them in the digital oscilloscope screen and marking with the two oscilloscope cursors. This phase difference arises between the induced *e.m.f.* in the ring and the exciting current in the main coil. The squared sine of this phase is the one that appears in Eqs. (11) and (12) of *F(z,t)* and its average *F(z)*, respectively. This phase difference appears mentioned and discussed in previous works on the jumping ring [2, 3]

**(b) Comparing the electric current in the jumping ring with the exciting current in the coil**

The exciting current in the main coil $i_{coil}$ (dashed curve) and the actual induced current $i_{ring}$ that appears in the ring (continuous curve) are plotted against time in Fig. 4, for a useful comparison. To obtain the current in the ring we numerically integrated the induced *e.m.f.* that appears in the pick-up coils (generated by the magnetic field of the ring itself, and shown in the lower trace of Fig. 3) and the results we plotted in the figure in kA units (see also Appendix B). The jumping ring (small nominal resistance of *0.070 mΩ* ) is seen to carry currents that can reach a thousand amperes, while the exciting current amplitude in the main coil reaches less than 10 amperes.

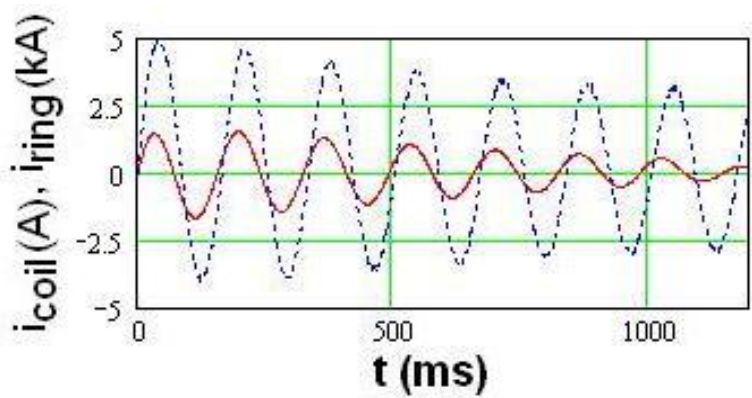

**Fig. 4** Exciting electric current in the main coil $i_{coil}$ (dashed curve) compared with the actual current induced in the ring $i_{ring}$ (note the kA units for this current) and plotted against time.

Alternatively, we also measured the integral of the *e.m.f.* induced in the ring (lower trace of Fig. 3) by connecting a passive *RC*-integrator to the two pick-up coils terminals. It is well-known [10, 11] that provided the condition $\omega RC >> 1$ is fulfilled the output voltage of this kind of electronic integrator is proportional to $i_{ring}(t)$. In our case we connected a resistor of *R= 20 k0hm, and a condenser of C= 1 µF* to the terminals of the pick-up coils.

### (c) Vertical position of the flying ring as a function of time during and after the thrust

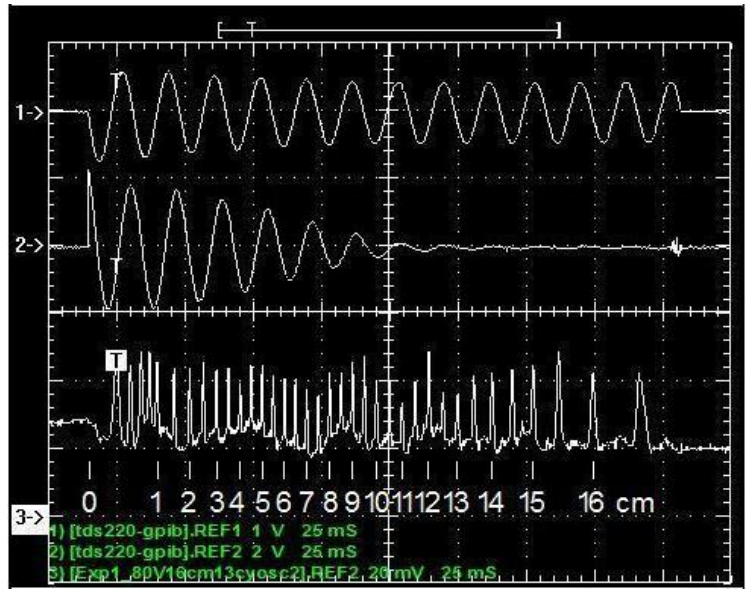

**Fig. 5** Oscilloscope trace of the vertical position of the ring (lower trace) as it travels upward (for comparison we copied Fig. 3 above the trace). Note the superimposed cm scale below the lower trace that allows one to measure the actual ring position during its flight (note that there are 5 data points in just the first centimeter of ring travel).

This registered experimental data on the ring position as it travels upward appears in the lower oscilloscope trace of Fig. 5 (note the superimposed cm scale). It is the signal generated in the phototransistor-diode circuit we use to detect the varying intensity of the transmitted beam of light by the ruled transparent strip as it moves upward while attached to the travelling ring. And, in Fig. 6 we have plotted this measured vertical position of the ring as a function of time; it is a truly useful and interesting figure. The two continuous plots in Fig. 6 are predicted ring positions using our theoretically model (Section 3), and correspond to two different numbers of exciting cycles, namely 13 and 3 cycles of applied AC voltage to the main coil (with an 80 V amplitude).

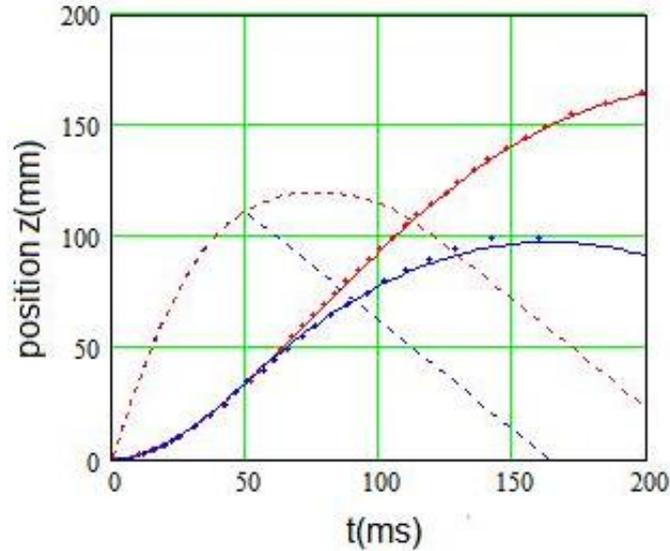

**Fig. 6** Position of the flying ring against time, for 3 cycles (lower continuous curve) and 13 cycles (upper continuous curve) of applied sinusoidal voltage to the main coil, and as predicted by our theory. The dots represent actual experimental data: note the good agreement with theory. The two dashed curves represent the speed of the ring obtained taking the derivatives of the continuous lines (longer speed dashed curve for the longer 13 cycles thrust).

As expected for a thrust lasting only 3 cycles (equivalent to 50 ms) of excitation voltage our ring fast rises up to a maximum height of about 10 cm in about 170 ms, in comparison it reaches about 17 cm when thrust during 13 cycles (upper continuous curve) in only 200 ms. The sets of large number of dots on top of (or very close to) both continuous curves in Fig. 6 are our actual experimental data. Note the good agreement between our experimental results and the predictions of our analytical model. It is remarkable that for both 3 cycles and 13 cycles, of applied voltage to the main coil, the theory predicted ring positions, as well as the two sets of experimental points are completely coincident in a single arc of curve (just above the first 50 ms in the time axis). After 50 ms the two theoretically obtained curves begin to depart gradually as expected, and again the experimental points closely follow the departure.

In Fig. 6 we have also plotted (dashed curves) the vertical speed $v(t)$ of the ring for both the 3 and 13 cycles thrusts. Again, it is remarkable to see that in both cases the ring vertical speeds are coincident just up to 50 ms (period of time of the shorter excitation). Afterwards, and for the 3 cycles case the vertical speed of the ring diminishes at a constant rate, because by then the ring is on the grip of the force of gravity (constant acceleration) and it carries on ascending up with decreasing speed to reach maximum height when its speed falls to zero ($v=0$ at about $t=170\ ms$). The velocity curve for the 13-cycles thrust case is the upper dashed curve in Fig. 6. It may be seen that contrary to the 3-cycles thrust case, the ring speed continues increasing up to about 75 ms when it reaches its maximum.

**(d) Maximum height of the ring as a function of the number of excitation cycles**

One of the often mentioned issues of the jumping ring experiment is the relation between the maximum height $h_{max}$ reached by the ring and the number of cycles of applied impulse. In Fig. 7 we have plotted our experimental results for three different values of the applied voltage to the main coil, namely 80 V (circles), 100 V (diamonds) and 118 V (crosses).

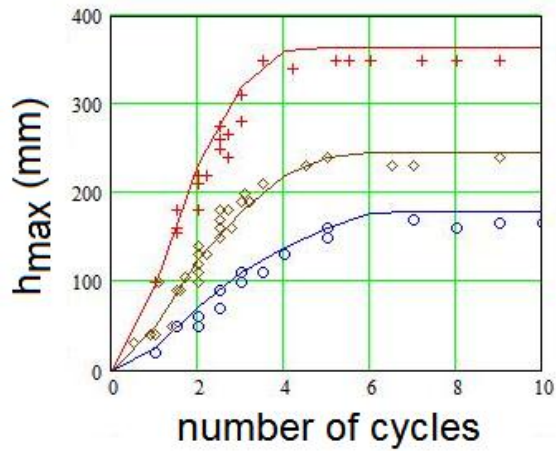

**Fig. 7** Maximum height reached by the ring as a function of the number of cycles of applied sinusoidal voltage to the main coil (continuous curves plotted using our theoretical model). The data points plotted correspond to voltage amplitudes of 118 V (crosses), 100 V (diamonds) and 80 V (circles) applied to the main coil. Note the good agreement between our theory and our experiments.

The three continuous curves in Fig. 7 were plotted using our theoretical model above (Section 2). Note the good agreement between our predictions and the experimental data. For any given constant amplitude of sinusoidal voltage applied to the main coil the maximum height reached by the ring increases almost linearly with the number of cycles of thrust to become constant later on (e.g. after 4 cycles for the 118 V curve). The explanation to this fact is quite simple: the ring has already abandoned the region where the electromagnet magnetic field exists (this occurs at about 36 cm above the core in our set-up, in the case of the upper curve).

### (f) The magnetic force on the ring as a function of time

Fig. 8 shows an important dependence to consider if one is to explain the magnetic force on the jumping ring: The instantaneous oscillating magnetic force (see Eq. (1)) on the ring appears plotted against time, as well as the average magnetic force on it (the decaying function, see Eq. (12)).

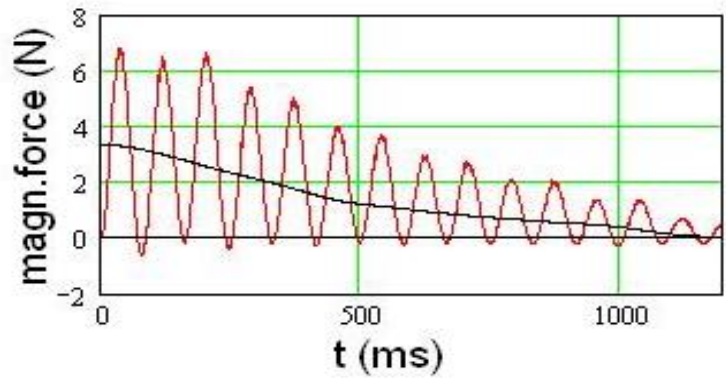

**Fig. 8** Magnetic force on the jumping ring as a function of time. The oscillating curve (at twice the frequency of the AC voltage applied to the main coil) represents the instantaneous force on the ring while the decaying curve represents the average force.

It may be seen that even for a time equal to a fraction of a cycle (say a quarter of a cycle) of the applied sinusoidal voltage to the electro-magnet there is going to exist a non negligible impulse applied to the ring. Also note in the figure that the oscillating force goes below zero, for small periods of time. The instantaneous force on the ring being the proportional to the product of the current in the ring times the excitation current in the coil and was therefore measured by us experimentally (see the currents plotted in Fig. 4): This instantaneous magnetic force is positive most of the time, yet it becomes negative −a fact related to the phase difference $\phi(z)$ between the two currents− for time intervals of duration $(\pi/2 - \phi(z))/\omega$.

### V. THE AUXILIARY FUNCTIONS

#### (a) The axial and radial functions $b_\rho(z)$, $b_z(z)$ of the electro-magnet field

In Fig. 9 we have plotted the measured auxiliary radial function $b_\rho(z)$ (circles) at the ring, and the auxiliary axial component $b_z(z)$ (crosses), of the electro-magnet magnetic field against the vertical coordinate *z*. The measurements of $b_\rho(z)$ were done with the *x-y-z* magnetometer, and the $b_z(z)$ with the small auxiliary (1 mm high) pick-up coil (mentioned at just the end of Section III) which we coaxially slide up along the core, and above it. The two continuous lines plotted are just fittings to the data. Note that the field auxiliary function $b_z(z)$ decay almost linearly, while the other $b_\rho(z)$ decays abruptly after 11 cm height, in coincidence with the protruding length of the core (above the main coil).

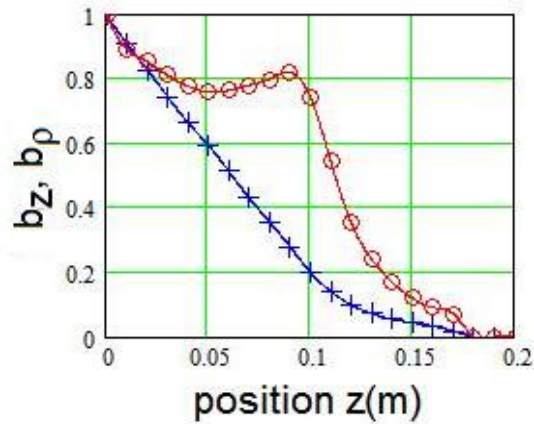

**Fig. 9** Auxiliary functions $b_z(z)$ (crosses) and $b_\rho(z)$ (circles) experimentally measured. The continuous lines are cubic-spline interpolation fittings to the data.

b) **The auxiliary function *l(z)* of the coil inductance and phase difference between the coil current and the current induced in the ring**

It was mentioned above (Section IV (a)) that the effective inductance *L'* of the main coil does not remain constant as the ring travels upward (see also Appendix A). Therefore, *L'* was written (see Eq. 12) as the product $L'(z) = L_0\, l(z)$.

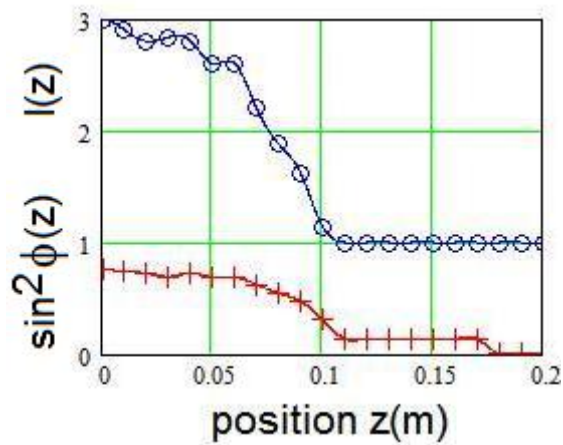

**Fig. 10** Measured auxiliary function *l(z)* (upper curve) and the function *sin²(ϕ)* (lower curve).

Figure 10 shows (upper curve) the measured auxiliary function l*(z)*. It may be seen this function decays from its initial low value of about 3 (ring resting on top of the coil) to 1 at about *z=11* cm above the coil, from there onwards it remains constant. The continuous curve in the upper curve of the figure was obtained using a cubic-spline fitting to the data. The lower curve in Fig. 10 shows instead the squared-sine of the phase difference *ϕ* between the induced *e.m.f.* in the ring and its own induced electrical current: plotted as the *sin²(ϕ)* against the vertical position *z*. The two curves in Fig. 10 are not independent from each other, and are of course related through $\omega L_0 \, l(z) = R \tan(\phi(z))$. This phase difference *ϕ* was measured in a preliminary experiment, in which the ring was positioned at fixed heights z, and each time we recorded both the sinusoidal-varying current in the coil, and the sinusoidal-varying *e.m.f.* induced in the pick-up coils in the digital dual-trace oscilloscope. As already mentioned the comparison of the two traces in Fig 3 also gives us this important phase difference *ϕ*.

## VI.    DISCUSSION AND CONCLUSIONS

Unveiling the physics behind the flight of the conductive jumping ring in the Thomson experiment is not a straightforward task because the electromagnetic variables of the *ring-electromagnet system* implicitly depend on the ring motion itself. This motion takes place in a divergent magnetic field, and thus we faced a true electro-dynamics problem. The reader would find that some of the observables of the ring motion have been presented in the pertinent literature without proper experimental support *e.g.* the maximum height reached by the ring, the instantaneous magnetic force on the ring, or the electrical current in the ring. In the present work we have applied two key ideas to unveil the physics of the ring motion: Firstly, in our theoretical model we separated the time and the vertical coordinate dependences of the main variables, introducing auxiliary functions that simplified our mathematical treatment, functions that we later managed to measure with relative ease in the laboratory. Secondly, a couple of long pick-up coils connected in opposition, and wound over the electromagnet core, allowed us to measure the magnetic field created by the moving magnet in spite of the presence of the much stronger field of the electro-magnet. Our approach allowed us to write the motion differential equation of the system, an equation that demanded a numerical solution using a well-known algorithm. In the motion equation we introduced the expression for the averaged force on the magnet previously derived in our model. We carried out a set of key experiments that have revealed for instance how the ring is accelerated upward by the thrust of the excited electromagnet, to end up in a ballistic flight upward (see Fig. 6). Both, the time dependence and the ring position dependence of the magnetic force on the ring have been theoretically derived using the model developed in the work, and then confirmed experimentally with good accuracy. In our results we included a plot (Fig. 8) of the measured instantaneous force on the ring, plot that was compared with the averaged force plot given by the theory, a comparison not found in the published literature. We have shown how the mutual inductance varies with the ring height and its concomitant effect on the main coil excitation current (Appendix A). The dependence of the maximum height reached by the ring upon the number of excitation cycles for different excitation currents was also presented (Fig. 10) for the first time. Our experiments were done with equipment ordinarily found in a physics laboratory, e.g. the digital dual-trace digital oscilloscope, and the low-cost set-up we used was assembled using either home-made parts or low-cost standard teaching laboratory equipment. Both, our model and our experimental procedures can now be extended, for instance to study the ring flight when a strong pulse of current be applied to the electromagnet. Hopefully, this work will become useful for a senior physics teaching laboratory, and also found useful to physics teachers at all levels, as well as to lecturers that present the Thomson jumping ring.

**APPENDIX A. The variation of the main coil current as a function of the ring vertical position: the auxiliary function *I(z)***

To that effect we present a transformer theory-like of the standard jumping ring set-up. The coil of the electro-magnet is considered to be the primary circuit (resistance $R_1$ and nominal inductance $L_1$) while the ring is the secondary circuit of the transformer (resistance $R_2$, and nominal inductance $L_2$). The Variac excites the primary with a voltage signal *V* of angular frequency $\omega$ and a current $i_1$ appears in it. If *M* denotes the mutual inductance between ring and coil, Kirchhoff Voltage Law for the coil gives

$$j\omega L_1 i_1 + R_1 i_1 = V + j\omega M i_2, \tag{A1}$$

while for the secondary circuit (no external voltage applied in it) we get

$$j\omega L_2 i_2 + R_2 i_2 = j\omega M i_1 \tag{A2}$$

Therefore the current $i_2$ in the ring is given by

$$i_2 = \frac{j\omega M}{R_2 + j\omega L_2} i_1. \tag{A3}$$

Replacing this expression into Eq. (A1) we get an equation for the current $i_1$ in the coil

$$i_1(j\omega L_1 + R_1) = V - \frac{(\omega M)^2}{R_2 + j\omega L_2} i_1,$$

or $\quad \left\{ R_1 + \frac{(\omega M)^2 R_2}{R_2^2 + \omega^2 L_2^2} + j\omega L_1 \left[ 1 - \frac{(\omega M)^2 L_2}{L_1(R_2^2 + \omega^2 L_2^2)} \right] \right\} i_1 = V \tag{A4}$

The resistance $R_2$ (of order 0.01 mΩ) of the ring is always such that $R_2 << \omega L_2$ and thus it is safe to write

$$\left\{ R_1 + \frac{M^2 R_2}{L_2^2} + j\omega L_1 \left[ 1 - \frac{M^2}{L_1 L_2} \right] \right\} i_1 = V, \tag{A5}$$

and again note that in this equation $\frac{M^2 R_2}{L_2^2} \ll R_1$, henceforth it is also safe to write

$$\left\{ R_1 + j\omega L_1 \left[ 1 - \frac{M^2}{L_1 L_2} \right] \right\} i_1 = V \tag{A6}$$

thus we can finally write the main coil current in terms of its own parameters

$$i_1 = \frac{V}{R_1 + j\omega L_1'} \tag{A7}$$

But a close examination of the last two equations reveals that the actual inductance $L_1'$ of the main coil (the primary circuit) in the jumping ring system is given by

$$L_1' = L_1 \left( 1 - \frac{M^2}{L_1 L_2} \right). \tag{A8}$$

And as the ring is thrust and travels upward the coupling mutual inductance $M$ diminishes, thenceforth the coil inductance $L_1'$ increases, and the current in the coil decreases as can be observed in the upper oscilloscope trace in Fig. 3. In Fig. A-1 we have plotted the results of measuring the coil current, and our plot of the auxiliary function *I(z)*. The mutual inductance *M* is a function of *z* and a relation to evaluate it is given in Appendix B, just below.

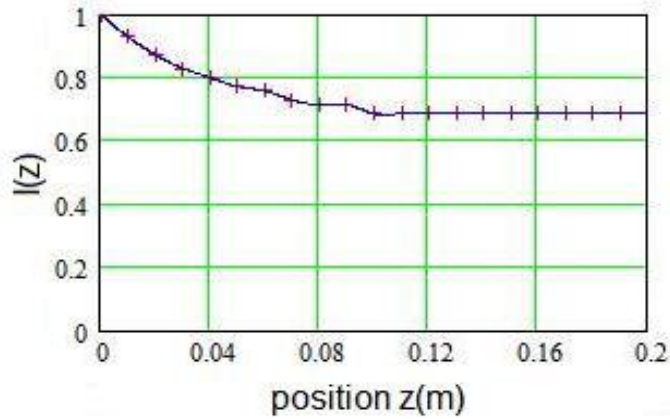

**Fig. A-1** Auxiliary function *I(z)* related to the decreasing mutual inductance between the main coil (primary) and ring (secondary). The continuous line is a cubic-spline fitting to the data.

**APPENDIX B  The *e.m.f.* induced in the long pick-up coils and the current $i_{ring}$ induced in the ring**

In this appendix we derive an additional relation between the *e.m.f.* induced in the long pick-up coils and the current $i_{ring}$ induced in the ring (of mean radius *a*) by the excited electromagnet. We begin considering the field created by the ring itself along its *z*-axis. It is given by a well-known textbook 6] relation

$$B_{z,ring}(z) = \frac{\mu_0 \, \mu_r(z) \, i_{ring} a^2}{2(a^2+z^2)^{3/2}} \tag{B1}$$

where $\mu_0$ and $\mu_r(z)$ are in our case the vacuum magnetic permeability and the relative permeability of the electro-magnet core, respectively.

Let *N* and *l* be the total number of turns and the length of the long pick-up coils (Fig. 1) that surround the core. The magnetic flux that appears in the coils, is given by

$$\Phi = \int b^2 \frac{\mu_0 \, \mu_r(z) \, i_{ring} a^2}{2(a^2+z^2)^{3/2}} dN, \tag{B2}$$

where $b^2$ is the core cross-section area, *dN=(N/l)dz*, and the integral is to be evaluated from *z= −∞* to *z= +∞*. Assuming now that the relative permeability of the core is a slowly varying function of *z  i.e.  $\mu_r(z) \approx \mu_r$* we get

$$\Phi = b^2 \frac{\mu_0\, \mu_r\, i_{ring} a^2 N}{2\, l} \left[\frac{z}{a^2(a^2+z^2)^{1/2}}\right]_{-\infty}^{+\infty} = \frac{b^2 \mu_0\, \mu_r N\, i_{ring}}{l} \tag{B3}$$

Finally, the amplitude of the induced *e.m.f.* that appears in the long pick-up coils is given by

$$\varepsilon_i = -\frac{d\Phi}{dt} = -\frac{N}{l} b^2 \mu_0\, \mu_r\, \frac{d}{dt} i_{ring}, \tag{B4}$$

and then the current in the ring can be obtained by integrating the *e.m.f.* that appears in the pick-up coils (as claimed above, Section IV of the main text).

Note that the varying mutual inductance between the travelling ring and the electromagnet is given by $M = \Phi/i_{ring}$. Required in Appendix A it can be obtained from Eq. (B3) by just changing the limits of the integral in that equation: from an initial z (the vertical coordinate of a given ring position) to $(z + l)$, where $l$ is now the length of the main coil. We then get the *z*-dependent function of the mutual induction $M$

$$M(z) = \pi a^2 \frac{\mu_0\, \mu_r N}{l} \left[\frac{z+l}{(a^2+(z+l)^2)^{1/2}} - \frac{z}{(a^2+z^2)^{1/2}}\right] \tag{B5}$$